\documentclass[acmtog]{acmart}

\AtBeginDocument{%
  \providecommand\BibTeX{{%
    Bib\TeX}}}

\usepackage{multirow}
\usepackage{makecell}
\usepackage{pifont}
\usepackage{caption}
\usepackage{subcaption}
\usepackage{graphicx}
\usepackage[export]{adjustbox}
\usepackage{lipsum}

\usepackage{amssymb}
\usepackage{amsmath}
\usepackage{threeparttable}
\usepackage{multirow}
\usepackage{subcaption}
\usepackage{algorithm}
\usepackage[noend]{algpseudocode}

\usepackage{color,soul}

\usepackage{lineno}
\usepackage{enumerate}
\usepackage{array}
\usepackage{booktabs}
\usepackage{etoolbox}
\usepackage{makecell}
\usepackage{mathtools}
\usepackage{wrapfig}
\usepackage{adjustbox}
\usepackage{lscape}
\usepackage{epstopdf}
\usepackage{multirow}
\usepackage{makecell}
\usepackage [adversary]{cryptocode}
\usepackage{latexsym}
\usepackage[misc]{ifsym}
\usepackage{pifont}
\newcommand{\cmark}{\ding{51}}%
\newcommand{\xmark}{\ding{55}}%
\usepackage{listings}
\usepackage{bm}
\usepackage{framed}
\usepackage{courier}
\usepackage{array}
\usepackage{tabularx}
\usepackage{xcolor,soul}
\usepackage{textcomp}
\usepackage{empheq}
\usepackage{subcaption}
\usepackage{float}
\usepackage{xfrac}
\usepackage{amssymb}
\usepackage{breqn}
\usepackage{xcolor}
\usepackage[T1]{fontenc}
\usepackage{svg}
\usepackage{relsize}
\usepackage{hyperref}
\usepackage{natbib}
\begin{document}

\title{Efficient and Privacy Preserving Group Signature for Federated Learning}


\author{Sneha Kanchan }
\email{snehakanchan159@soongsil.ac.kr}
\affiliation{\institution{School of Computer Science and Engineering, Soongsil University, Seoul}
\country{South Korea}}
\author{Jae Won Jang}
\email{jwon0524@soongsil.ac.kr}
\affiliation{\institution{School of Computer Science and Engineering, Soongsil University, Seoul}
\country{South Korea}}
% \ead{Author2 email}
\author{Jun Yong Yoon}
\email{wnsdyd1124@soongsil.ac.kr}
\affiliation{\institution{School of Computer Science and Engineering, Soongsil University, Seoul}
\country{South Korea}}
\author{Bong Jun Choi}
\email{davidchoi@soongsil.ac.kr}
\affiliation{\institution{School of Computer Science and Engineering, Soongsil University, Seoul}
\country{South Korea}}
\renewcommand{\shortauthors}{Trovato et al.}

\begin{abstract}
   Federated Learning (FL) is a Machine Learning (ML) technique that aims to reduce the threats to user data privacy. Training is done using the raw data on the users' device, called clients, and only the training results, called gradients, are sent to the server to be aggregated and generate an updated model. However, we cannot assume that the server can be trusted with private information, such as metadata related to the owner or source of the data. So, hiding the client information from the server helps reduce privacy-related attacks. Therefore, the privacy of the client's identity, along with the privacy of the client's data, is necessary to make such attacks more difficult. This paper proposes an efficient and privacy-preserving protocol for FL based on group signature. A new group signature for federated learning, called GSFL, is designed to not only protect the privacy of the client's data and identity but also significantly reduce the computation and communication costs considering the iterative process of federated learning. We show that GSFL outperforms existing approaches in terms of computation, communication, and signaling costs. Also, we show that the proposed protocol can handle various security attacks in the federated learning environment. Moreover, we provide security proof of our protocol using a formal security verification tool.
\end{abstract}

\begin{CCSXML}
<ccs2012>
 <concept>
  <concept_id>10010520.10010553.10010562</concept_id>
  <concept_desc>Computer systems organization~Embedded systems</concept_desc>
  <concept_significance>500</concept_significance>
 </concept>
 <concept>
  <concept_id>10010520.10010575.10010755</concept_id>
  <concept_desc>Computer systems organization~Redundancy</concept_desc>
  <concept_significance>300</concept_significance>
 </concept>
 <concept>
  <concept_id>10010520.10010553.10010554</concept_id>
  <concept_desc>Computer systems organization~Robotics</concept_desc>
  <concept_significance>100</concept_significance>
 </concept>
 <concept>
  <concept_id>10003033.10003083.10003095</concept_id>
  <concept_desc>Networks~Network reliability</concept_desc>
  <concept_significance>100</concept_significance>
 </concept>
</ccs2012>
\end{CCSXML}

\ccsdesc[500]{Computer systems organization~Embedded systems}
\ccsdesc[300]{Computer systems organization~Trust}
\ccsdesc{Computer systems organization~Privacy}
\ccsdesc[100]{Networks~Network reliability}

\keywords{Federated Learning, Group Signature, Privacy Preservation, Authentication, Efficiency, Adversarial Server}

\maketitle

\section{Introduction}
Federated Learning (FL) is a machine learning method ensuring that raw data is not distributed to other devices. The motivation behind FL is to secure users' sensitive information. For example, uploading real-time images increases the risk of revealing personal information to the server or eavesdroppers. The data breach has become a serious threat that even if organizations spend a massive budget on securing their data, they are still getting attacked. In 2020, the investment to secure companies' data grew to 53 billion. However, 30 billion records were compromised \cite{chris}. It shows how serious and important it has become to secure our data. It is often illegal to share data from one place to another, even from one country to another. Therefore, FL will be essential in sharing machine learning models without compromising privacy. 

% Users might be willing to share their data but not their details. For example, suppose Google wants to train an object detection model and needs lots of data for the training purpose.

\begin{figure}
    % \begin{subfigure}{.5\textwidth}
    %     \centering
    %     \includegraphics[width=.95\linewidth]{server is compromised.png}
    %     \caption{When the server is compromised}
    %     \label{1a}
    % \end{subfigure}
    % \begin{subfigure}{.5\textwidth}
    %     \centering
    %     \includegraphics[width=.95\linewidth]{client is compromised.png}
    %     \caption{When a client is compromised}
    %     \label{1b}
    % \end{subfigure}
    % \begin{subfigure}{.45\textwidth}
        \centering
        \includegraphics[width=.9\linewidth]{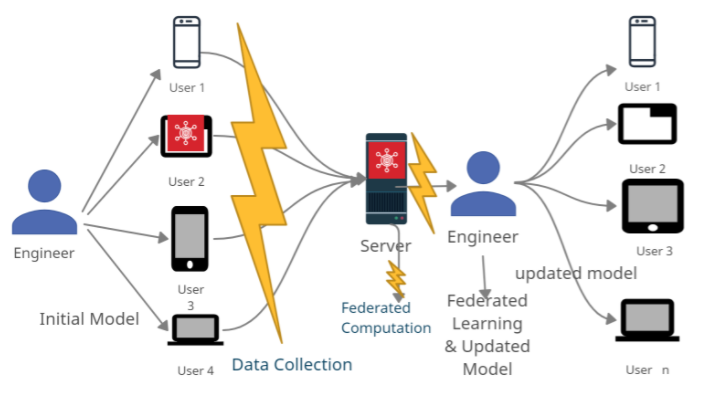}
        \caption{Scenario when client/clients and the server are compromised}
        \label{1c}
    % \end{subfigure}
    % \begin{subfigure}{.49\textwidth}
        % \centering
        % \includegraphics[width=.95\linewidth]{engineer and server are compromised.png}
        % \caption{When various network entities in FL network is compromised}
        % \label{1d}
    % \end{subfigure}
    % \caption{Various security risks of federated learning when network entities are compromised: a) As the server is the major resource to collect data from the clients, it can affect most of the data transmitted at different levels, b) The compromised client can try to attack other clients if it communicates with them, c) They can collude to launch a larger scale attack or hide their attacks, d) Client's data and updated model can be affected as well, which can be used to gain important information in upcoming iterations.}
    \label{fig1}
\end{figure}

The clients in a FL network participate in updating the system model by sending their local model updates to a central system, say a server. Then, the server can finalize the global model to be deployed. During this process, the clients do not send their raw data. They send only locally trained gradients. Next, the server collects gradients from all the clients and aggregates those to generate an updated global model. Still, there are many scenarios at various stages of the FL process where network entities, including the server, are compromised, requiring additional measures to preserve privacy even if the data is not being sent in raw form, as shown in Figure \ref{fig1}. The communication process mainly involves the participating clients and the server. If any of the entities in the network is compromised, the secrecy of this technique is at stake. In that case, the compromised entity can attempt to infer the sensitive information or the source of information from the received gradients based on the sender, revealing the sender identity \cite{hu} and increasing the risk of inference attacks. Hence, it is crucial to secure the privacy of the sender and the information even if only the updated gradient is being shared. Moreover, since FL is a decentralized network, entities of the network can be active or passive attackers themselves \cite{aslam}. 

\subsection{Related Work}

The conventional privacy-preserving techniques for FL, their contributions, and their limitations are discussed below:

\subsubsection{\textbf{Masking Based}}

Bonawitz et al. \cite{Bonawitz} proposed a double masking scheme in which messages are shared with servers after double masking it. Each user shares double sets of secret keys with every participant. After adding their share of secret keys to the message, all participants send messages to the server. Every secret key gets canceled out when a certain number of secret keys are combined. Hence, actual data is hidden, and the method effectively protects privacy. However, the computation cost is exponentially high because the secret keys must be distributed to each client for each message.
In 2020, Li, Yong et al. \cite{yong} proposed a single masking FL framework based on the chained secure multiparty computing (SMC) technique (chain-PPFL). They used a chained-communication mechanism to transfer the masked information between participants with a serial chain frame. Their experimental result shows they can achieve competent accuracy and convergence rate while providing privacy preservation for FL. However, their costs are higher than many existing techniques.
In 2021, Ang Li et al. \cite{Ang} addressed the issue of heterogeneous data in separate clients and the bandwidth limitation of a mobile device. They proposed the FedMask FL framework, where only heterogeneous binary masks are communicated between the server and the end devices. The clients do not learn the global model. Still, they learn from a personalized and structured sparse DNN model composed by applying the learned binary mask to the fixed parameters (frozen parameters) of the local model. However, the model has high complexity.

\subsubsection{\textbf{Homomorphic Encryption (HE) Based}}

Wan et al. \cite{wan} proposed a Verifiable Federated Matrix Factorization (VPFedMF) technique, which highlights the shortcomings of earlier proposed additive homomorphic encryption and FedMF, as these had significant performance overheads, weaker security model, and non-assurance of computation integrity of model-aggregation. In comparison, VPFedMF's masking-based lightweight and secure aggregation provide cryptographic guarantees on the confidentiality of gradient and user-side verification for the correctness of the aggregated model.

\subsubsection{\textbf{Group Based}}

In 2018, Hartmann \cite{Hartmann} proposed sampling-based groups in FL in which users were divided into various groups related to their physical traits like location, region, language, etc. The tasks of the members are divided by their groups, where different groups execute independent tasks. Differential privacy (DP) is used for maintaining privacy. However, the computation cost of the algorithm is high, and it is not scalable in more extensive networks.

\subsubsection{\textbf{Blockchain Based}}

In 2020, Kim et al. \cite{kim} proposed a blockchain FL (BlockFL), in which blockchain was used to exchange information between clients. Their protocol is based upon on-device machine learning, and there is no centralized data for training. This model focuses on exchanging global models with their associated minor, and they receive awards based on the trained samples transmitted. However, storage and speed limitations are major bottlenecks with on-device training. 
Recently, in 2022, Zhu et al. \cite{Zhu} proposed a double-masked blockchain-based FL technique, in which they claimed to provide better privacy with lower communication costs between blockchain and end-users. However, it has high computation costs for computing double masks. Also, the problem of sharing secrets with other participants and user dropouts remains.
Li et al. \cite{yijing} presented a FL-based privacy preservation technique in autonomous driving where the original data is stored in local vehicles, and only updated information is shared with the server. Based on blockchain, drivers are authenticated with zero-knowledge proof. They used homomorphic encryption in their traceable identity-based approach because the server cannot be trusted completely. 
    
%Their technique is to protect the privacy of data and its source. \textcolor{red}{[David: Any drawbacks?]}

\subsection{Limitations of Existing Approaches}

\subsubsection{\textbf{Masking Based}} 

The cost of communication is relatively high. Each client needs to communicate with other clients in the network to share the secret keys, making the complexity in polynomial time, which is highly inefficient. Also, the client needs to trust another client with their secret keys, which is risky because nodes in a decentralized wireless system cannot be trusted. The situation worsens when the server and a client (or few clients) conspire together to extract information about a particular client (Figure \ref{1c}). If the server can acquire a particular number of secret keys from the compromised clients, it can reveal the actual data sent by a client. Moreover, the list of clients participating in the FL process is disclosed publicly so that clients may know the participating clients to share their secret keys. However, the list is a piece of additional information for the attacker, who may target a particular client from the list.

% In the traditional masking approach, a client secures its data by adding secret keys. Each secret key has $n$ parts, where combining all parts results in null. When a client sends a message, it adds one secret part to its message. The remaining parts are shared with other clients who take part in communication. The clients bind their messages with all the secret keys received. Hence, each client's message has $n$ additional parts for secret keys, the first part for its secret key, and the rest $(n-1$) for other $(n-1)$ clients' secret parts. When the server receives these messages, it aggregates them, and all secret keys cancel each other. The gradients are aggregated, and the server does not know the individual gradients or their owner; thus, privacy is preserved. However, there are some limitations to this approach, as summarized below:h

\subsubsection{\textbf{Homomorphic Encryption based}}

Clients send encrypted data to a server, aggregating the received encrypted data without decrypting it. HE can reduce the model inversion or data leakage risks even if the server is malicious. However, it has a high complexity making it challenging to be implemented in practice. The server might need some privacy-relevant data to aggregate the message. Also, the output obtained from HE may contain more noise than the original data, so the accuracy of the model trained using these data is likely to be reduced.
% That is where differential privacy methods can be useful in addition to HE. However, using these algorithms increases computational complexity by a considerable margin.

\subsubsection{\textbf{Differential Privacy Based}}

DP ensures that the outcome does not reveal whether an individual's data has been included in the database or not \cite{Wei}. It helps to secure the privacy of individual data as the behavior of the output data hardly changes when a single individual joins or leaves the dataset. However, if data is very much diversified or we have to retrieve data from a vast dataset, DP adds noise to data in each step, reducing accuracy. 
 
\subsubsection{\textbf{Blockchain Based}}

Blockchain is well-suited technology to protect the distributed data in FL. However, there is redundant data on devices and in different rounds, which makes it inefficient for the iterative process of FL. Because of its complexity and storage requirements, scalability is a challenge. 

\subsection{Our Approach}

We are motivated to implement privacy in FL using a less complex, more efficient, and secure technology to solve the abovementioned issues. Hence, we propose a privacy-preserving Group Signature scheme for FL (GSFL). A group signature (GS) verifies a group instead of a particular client. There can be multiple groups, and each group consists of an admin. The admin is assumed to be a trusted entity in the network, assigning the keys to each group member to generate the signature. No one other than the admin knows the signer. Since the server cannot infer the particular identity from the received information, the risk of inference attacks is reduced because there is no linking between the clients and their gradients. GSFL has several advantages over the existing privacy-preserving techniques for FL, as explained below. 

%\textcolor{red}{[David: I think we are still directly comparing with HE and DP? I thought HE and DP provide data privacy. In contrast, GS provides user identity privacy?]} \textcolor{blue}{Sneha: Because people ask for reasons why we are not using HE or DP, why not just protect data instead of users. We need to give a reason why protecting the identity is important instead of just data privacy.}: 

\begin{itemize}
    \item \textbf{There is no need to share keys with other clients as in masking}: A client communicates only with an admin and a server. It considerably saves communication costs.
    \item \textbf{GS does not require data to be aggregated in encrypted form as in HE}: The trained model is encrypted with the server's public keys; hence only the server can decrypt it. However, the server does not know the source. 
    \item \textbf{Even if the data is diversified, there is no need to add noise to the data as in DP}: Since the noise level is lower in the original trained data, the accuracy rate is also higher. 
    \item \textbf{Privacy preservation and authentication are done together, takes lesser storage}: The members in a group share the same signature, and hence, the server needs to store less number of signatures to authenticate the clients.
\end{itemize}

% Even though the sender of the information is unknown to the server, the signer is not anonymous. 
GS is fundamentally different from anonymity, as later, there is no way to trace back the sender of the anonymous data. Hence, the information received from anonymous clients cannot be trusted. However, in GS, clients authenticate their identity as valid group members, and the admin can trace it. Hence the information shared by the group member is verifiable, making it more trustworthy than anonymous data.

Implementing GS in FL changes the communication process, but it does not affect the learning process. Therefore, it can be used in combination with different FL models. However, the original GS contains several verification parameters, increasing the packet's size. The distributed learning process in FL incurs a high communication cost from multiple iterations involving thousands of clients. Hence, communication efficiency is one of the critical performance metrics. Therefore, we aim to provide a new group signature algorithm that is efficient for the iterative process of FL. 

% Instead of signing updated gradients with a personal signature, we sign with GS. 

\subsection{Our Contribution}
The contribution of our proposed work is summarized below:
\begin{itemize}
    \item \textbf{Design a new group signature tailored for FL}: The proposed GS protocol provides an efficient integration with the iterative process of FL.
    \item \textbf{Provide identity preservation of clients}: GS hides the identity of individual clients to provide enhanced privacy protection. 
    \item \textbf{Provide authentication of clients}: Clients are authenticated using a common GS.
    \item \textbf{Maintain the secrecy of network}: The security of the network is verified by a formal verification tool AVISPA (Automated Validation of Internet Security Protocols and Applications) \cite{AVISPA}. 
    \item \textbf{Provide enhanced security and efficiency}: Our protocol achieves significantly lower computation and communication costs than existing algorithms while protecting against a more comprehensive range of security attacks. 
\end{itemize}

% \subsection{Improvements from the Conference Version}
% The initial version of this work was presented at \cite{sneha}, in which we have integrated the privacy preservation feature of the group signature in the FL process. This latest version has a lower cost associated with the original version. A detailed comparison of original work, existing techniques, and our contribution have been given in this paper. In addition, security analysis and attacks possibility has also been discussed in detail.

\subsection{Paper Organization}
The remainder of the paper is organized as follows. Section \ref{prem} presents preliminaries and our system model. Section \ref{prop} presents the detail of our proposed protocol. A comprehensive security analysis is provided in Section \ref{attack}. The performance analysis of our proposed protocol compared with existing algorithms is provided in Section \ref{perf}. Finally, the conclusion of our paper is given in Section \ref{con}. The list of symbols and abbreviations used in the paper are presented in Table \ref{TabV}.

\begin{table*}[t]
\caption{List of symbols and abbreviations used}
\label{TabV}
\centering
%\resizebox{9.0cm}{3.0cm}
{
\begin{tabular} [t]{l l}
\hline
 Notations and Symbols  & Description 
\\ \hline
$A_i$ & Corresponding signing component of client $i$ \\
$B_1,B_2,B_3,B_4,B_5$ & Binding variables to compute hashed message \\
$C_1,C_2,C_3$ &  Tracing variables to trace the original signer \\
$(c_1,c_2,\sigma_{GS})$ & Set of encrypted messages \\
$C_{BV} $ & Cost of binding variables \\
$ \Tilde{C_{BV}}$ & Cost of verifying binding variables
% $C_{CH}$ & Cost of hashing function \\
\\
$C_{client-client}$ & Cost of client to client communication \\

$C_{SV}$ & Cost of signing variables \\
$C_{TV}$ & Cost of tracing variables \\

$C_{agg}$ & Cost of aggregation  \\

$C_{dec}$ & Cost of decryption \\
$C_{end}$ & Cost of encryption \\
$C_{ver}$ & Cost of verification  \\ 
$d_\alpha, d_\beta, d_x, d_{\theta_1}, d_{\theta_2}$ & Binding values to compute signing variables \\
$FL$ & Federated Learning \\
$\gamma$ & Private key of admin \\
$GID$ & Group ID \\
$GSFL$ & Group Signature-based Federated Learning \\
$h$ & Hash function \\
$LatestGS$ & Latest group signature \\
$m$ & Total number of clients in the network \\
$n$ & Total number of clients participating in FL \\
$s_1,s_2,s_3,s_4,s_5$ & Signing variables to compute final signature \\
$\sigma_{GS}$ / $GS$ & Group Signature \\
$t$& Number of iterations\\
$Upd$ & Updated gradients after local training \\
$u,v,a,b$ & Exponents chosen for signature \\
$x_i$ / $ID$ & Real identity of client $i$ \\

\hline
\end{tabular}
} 
\end{table*}

\section{Preliminaries and our system setup} \label{prem}

Federated Learning involves an iterative process where users collaboratively train machine learning models while protecting the privacy of user data, as implemented in Google Gboard \cite{Timothy}. End devices (clients) process their data locally, and instead of the raw data, they send only updates to the centralized system (server). After this, data from all the clients would be aggregated at the server to create an updated global model, and this model is again sent back to clients. The iterative process continues until an optimum global model is obtained.

% At first,  decentralized data is collected from selected clients. For example, our mobile keyboards are the end devices. We cannot expect that people will hit the correct keystroke every time, and if they hit the wrong keywords, it should be corrected by a smart keyboard. The aim of FL is auto-correction and predictions here.

% This aggregation is a continuous iterative process, and in each iteration, there is a change in participating clients.

%, and the model would be sent to the engineer. In the next phase, engineers or analysts will analyze the aggregated updates and categorize the features useful in further updates. 

\subsection{Privacy Preservation in Federated Learning}

FL aims to minimize the information shared by the clients. Although the server can only see the updates, if the server can identify the client of a particular update, it can guess the original data because it has some additional information based on the previous iterations or causes inference attacks \cite{hu}. Hence, it is vital to protect the client's identity along with data. This can be achieved in various ways, as listed below:
\begin{itemize}

    \item \textbf{Homomorphic Encryption}: HE hides the individual updates and enables the aggregation process on encrypted data, which leads to malleability \cite{manoj}. The server does not know which actual data was sent, and data is aggregated without revealing the original content. However, the server can add something to the encrypted data, which is very much disadvantageous and may affect the integrity of the data. Moreover, it is complex to perform computation on encrypted data, and the source may need to reveal part of their secret key, a significant threat in a wireless scenario. Hence, HE is not entirely suitable as it is very complex and may lead to various security threats.
    \item \textbf{Masking} \cite{Brendan}: It hides the data using masks. The server cannot unmask individual data from a separate device and can be unmasked only when combined with other devices' data. So, the server can only see combined data and does not know the source. However, the cost of sharing masks is very high, as it requires each client to communicate with others for mask sharing. 
    \item \textbf{Group Signature}: It hides the identity of the data owner. The server will not know which update is coming from which device \cite{david}. However, the updates are visible to the server.
\end{itemize}

\begin{figure*}[!htb]
    \centering
   % \includesvg[width=.9\textwidth]{Simple.svg}
    \includegraphics[width = .99 \textwidth]{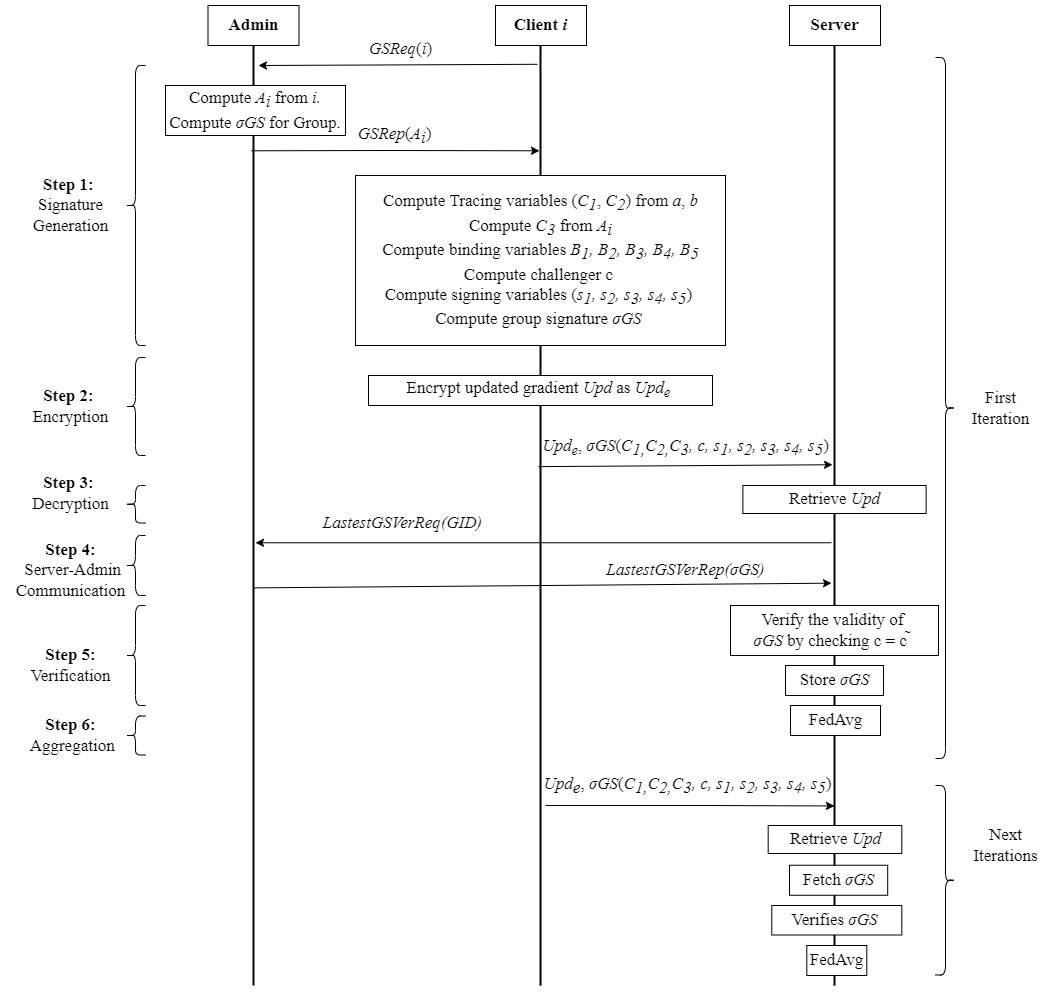}
    \caption{A simple integration of traditional group signature in federated learning (baseline)}
    \label{GStrad}
\end{figure*} 
We have chosen the group signature-based privacy preservation technique for our work. It can preserve the integrity of the original information to achieve high accuracy while having relatively low complexity compared to other works, especially when there are many clients and iterations in FL.

%However, data are not hidden from the server, either in encryption or masks. Data is encrypted before sending it out, but it is done mainly with the server's public key, making it easier for the server to decrypt the data and access it. However, the data sent in FL is never raw but is always locally processed. Only updates are sent to the server, and the server does not know the source of information. Hence, it cannot relate to the data obtained in previous iterations.

\subsection{Group Signature}

Group signature is proposed by \cite{david} in 1991, which shows how a person can be authenticated as a member of a particular group without revealing their identity. Only a valid member can sign the group signature. The receiver can always verify that the signature is from a valid group but can never verify the actual individual who signed the message. However, the issuer of the group signature can always find out the original signer of the message. Hence, a group signature assures privacy, authentication, and non-repudiation at the same time.

In our initial work \cite{sneha}, we have attempted to integrate GS in FL. As shown in Figure \ref{GStrad}, the admin issues $A_i$ to the client $i$, and this $A_i$ is used to create a group signature using other global parameters provided by the admin. Client $i$ signs its update using the group signature generated by it. It also sends a few parameters in the message along with the group signature, which is used for the verification process by the server. This process is repeated in each iteration of the FL process.

The use of a group signature assures that the privacy of the source of information is preserved even without the need for client-to-client communication. As shown in Figure \ref{GStrad}, the server cannot fetch $A_i$, but it can still verify the signature for authentication. This requires a much lower cost, and this process is significantly beneficial compared to traditional masking/double masking techniques. In this paper, we propose a new and improved group signature scheme for FL, where the computation and communication cost is significantly reduced for the iterative process. 

\subsection{Bilinear Pairing}

%The main application is to analyze cryptographic algorithms.

Bilinear pairing is the pairing of two multiplicative and isomorphic cyclic groups $G_{1}$ and $G_{2}$ to map into a third group $G_{T}$, with a mapping $e:G_{1}\times G_{2}\to G_{T}$. This map holds the bilinearity, non-degeneracy, and computability, as given in \cite{gsis}. Many identity-based encryption systems are based on these bilinear pairing concepts. The bilinear pairing can be used to solve a Decisional Diffie-Hellman (DDH) problem, which is the base of our identity-based cryptosystem. We have assumed $g_1$, $g_2$, and $g_t$ as the generator of the group $G_1$, $G_2$ and $G_T$, respectively, all of order a prime number $q$. We assume that Strong Diffie-Hellman (SDH) holds for $(G_1, G_2)$, and signature generation is based on a new Zero-Knowledge Proof of Knowledge (ZKPK) of the solution to an SDH problem \cite{Boneh}.

% \subsection{Client Selection by Server}

% The server can select the clients for a single update, which will continue for the entire iteration consisting of various iterations of gradient transmission. When the server announces that it is up for client selection, each valid member of the group sends their specification with the FL request. 

\section {Proposed Protocol} \label{prop}

\subsection{Problem Formulation}

We aim to improve the traditional group signature process considering the iterative process of FL. Traditionally, the group signature sends all the binding and tracing parameters in its signature so it can be recovered at the receiver side. It requires certain variables at the receiver side to verify that it is a valid group signature. Compared to the baseline protocol presented in \cite{sneha}, the proposed protocol improves the efficiency of group signature by avoiding repetitive transmission of some variables. We assume that a signature is valid for at least a session of the FL. We also assume that the admin is the trusted third party for all communication members, including clients and servers. Admin communicates with the server to share the latest valid group signature for the session. Let us call this signature as $\sigma_{GS}$. The tracing variable set $(C_1, C_2, C_3)$ contains three variables of 128-bit each, which is used to verify the signature at the receiver side. However, $(C_1, C_2)$ does not require any private key, separating one client from another. Hence, we can save computation and communication costs if the admin sends these variables directly to the server instead of the clients computing and sending them to the server.
\begin{figure*}[!htb]
    \centering
    %\includesvg[width=\textwidth]{GSFL.svg}
    \includegraphics[width=0.99\textwidth]{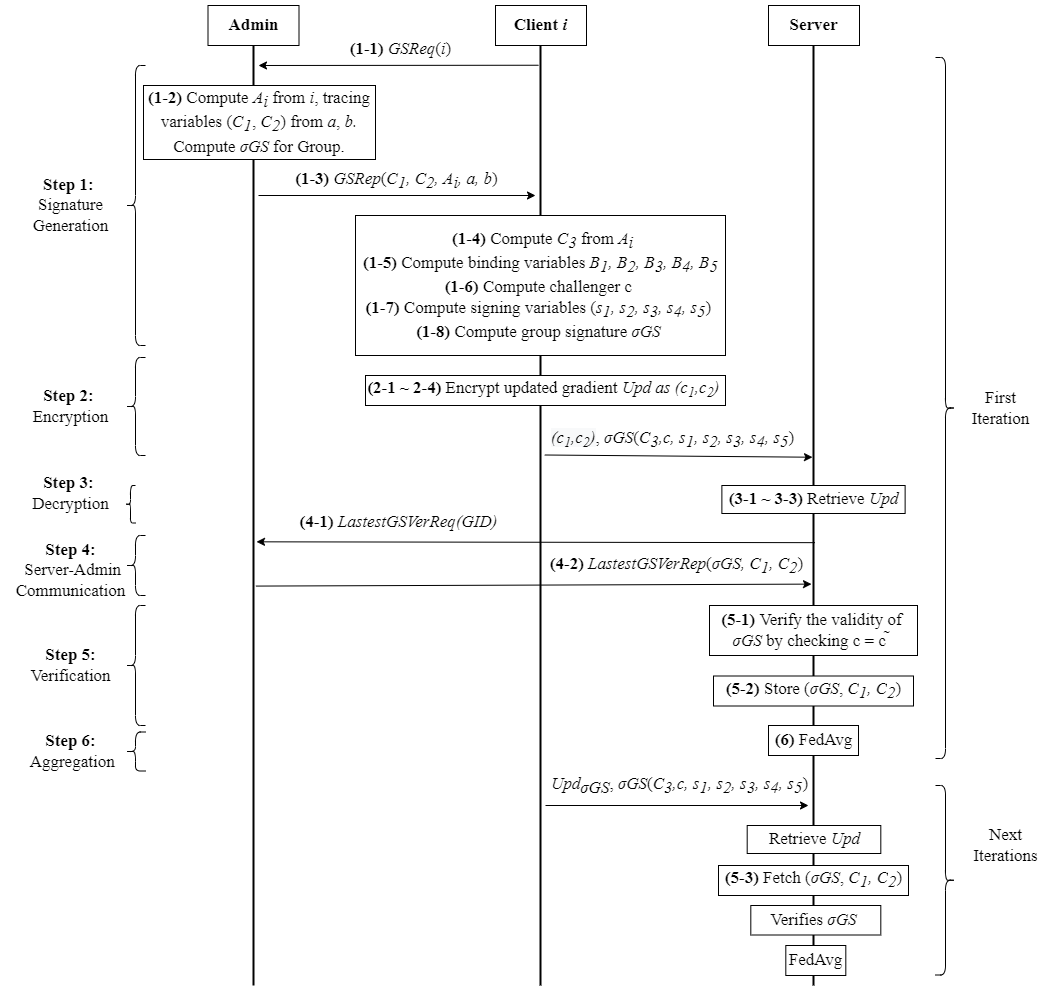}
    \caption{Proposed group signature generation and communication for iterative process of federated learning}.
    \label{FLGS}
\end{figure*}

\subsection{The Algorithm}

A novel efficient group signature for the iterative process of federated learning is presented in this subsection. The protocol diagram is shown in Figure \ref{FLGS}.

\subsubsection*{\textbf{Step 1: Generating Group Signature}}

Below are the steps to generate a group signature for a group. 

\begin{enumerate}
\item Client $i$ sends a Group Signature Request ($GSReq(i)$) to the Admin.

\item Admin computes Tracing Variables $C_1$ and $C_2$:

\hskip .2 in $C_1 \leftarrow u^a$, \hskip .2 in $C_2 \leftarrow v^b$,

where $a,b \in Z_p$ are the randomly chosen exponents selected by the admin. Admin also computes $A_i$ for the $i$-th client as:

\hskip .2 in $A_i \leftarrow g^{1/(\gamma + x_i)} \in G_1$.

At this stage, the admin also computes the group signature for the group, $\sigma_{GS}$, in the same way clients compute it in step 5 of this algorithm. This signature is stored on the admin for verification purposes.

\item  Admin sends a Group Signature Reply to the clients in the group generated as: 

\hskip .2 in $GSRep \leftarrow (C_1, C_2, A_i, a, b)$,

and sends the tracing variables and the group signature $(C_1, C_2, \sigma_{GS})$ to the server.

\item  The tracing variable generated at the client-side is computed as:

\hskip .2 in $C_3 \leftarrow Ah^{a+b}$, \hskip .2 in $\delta_1 \leftarrow xa$, \hskip .2 in $\delta_2 \leftarrow xb$. 

\item  Each client randomly selects the following five binding values, $d_\alpha, d_\beta, d_x, d_{\theta_1}$, and $d_{\theta_2}$. Each client computes its binding variables as:

\hskip .2 in $B_1 \leftarrow u^{d_ a}$, \hskip .45 in
$B_2 \leftarrow v^{d_b}$,

\hskip .2 in $B_3 \leftarrow \hat{e}(C_3,g_2)^{d_x} \hat{e}(h,w)^{- d_ a -d_ b} \hat{e}(h,g_2)^{- d_ {\delta _1} -d_ {\delta _2}}$,

\hskip .2 in $B_4 \leftarrow C_1^{d_x} u^{-d_{\delta _1}}$, \hskip .1 in $B_5 \leftarrow C_2^{d_x} v^{-d_{\delta _2}}$.

\item Each client computes its challenger $c$ as:

\hskip .2 in $c \leftarrow H(\omega,C_1,C_2,C_3,B_1,B_2,B_3,B_4,B_5) \in Z_p$.

\item  It computes signing variables from challenger:

\hskip .2 in $s_1 \leftarrow  d_a +ca$, \hskip .2 in $s_2 \leftarrow  d_b +cb$, \hskip .2 in $s_3 \leftarrow  d_x +cx$, 

\hskip .2 in $s_{4} \leftarrow  d_{\delta _1} +c{\delta _1}$,
\hskip .1 in $s_{5} \leftarrow  d_{\delta _2} +c{\delta _2}$.

\item Finally, the group signature is generated as:

\hskip .2 in $\sigma_{GS} \leftarrow (C_3,c,s_1,s_2,s_3,s_4,s_5)$.

\end{enumerate}

Compared with the baseline, the final signature consist of only $C_3$, instead of the original set $({C_1,C_2,C_3})$. Although the size of the algorithm is reduced, the overall unknown key size is reduced from 384-bits to 128-bits since the keys $({C_1, C_2})$ are known to the server and the same for all clients. Hence, to increase the unknown variable key size, we have increased the size of all three keys from 128-bits to 256-bits. The resulting size of the unknown key is still 256-bits compared to 384-bits earlier, but it is infeasible to break in real-time. Hence, it is safe to use. Please note that this signature is used for all iterations of the same session in Figure \ref{FLGS}. Hence, the signature generation step is only required once and not for subsequent iterations.

\subsubsection*{\textbf{Step 2: Encryption}}

Encryption is done to protect the message from outsiders. We used ElGamal encryption for our protocol, as given in \cite{moti}. The server publishes its public key set as $(G_T,q,g_t,h)$, where $h = g_t^x$ and $x$ is the private key of server. The message is encrypted with the server's public key, so the server can decrypt it to check the message signed with GS. The client encrypts message $M=Upd$ as follows:
\begin{enumerate}
    \item Chooses a random integer $y \in \{1,\ldots,q-1\}$.
    \item Compute $s = h^y$.
    \item Compute $c_1 = g_t^y$.
    \item Compute $c_2 = M\times s$.
    % where $N = Upd_{\sigma_{GS}}$ and $Upd$ is the updated gradient which needs to be sent. 
\end{enumerate}
% \begin{equation}
% c_1 = M_i^{s_{pk}} mod~n.
% \end{equation}
Then, the message set $(c_1,c_2,\sigma_{GS})$ is sent to the server. 

\subsubsection*{\textbf{Step 3: Decryption}}

Since the message is encrypted with the server's public key, it must be first decrypted at the server's side using the server's private key $x$. Server follows the following steps:
\begin{enumerate}
    \item Compute $s' = c_1^x$.
    \item Since $c_1 = g_t^y$, $s'$ = $(g_t^{y})^x = g_t^{xy} = h^y = s$.
    \item Compute $M = c_2s^{'-1}$. The inverse of $s'$ can be computed as $s^{'-1} = c_1^{q-x}$ \cite{wiki}.
\end{enumerate}

\subsubsection*{\textbf{Step 4: Server to Admin Communication}}

The server checks with the admin for the latest group signature of the group. 

\begin{enumerate}
    \item The server sends a request to the admin for the latest group signature of the group ($LastestGSVerReq(GID)$). 
    \item Admin replies to the server with the latest group signature version ($LastestGSVerRep(C1,C2)$) that contains $(\sigma_{GS}, C_1, C_2)$, where $\sigma_{GS}$ denotes the signature of that particular group and ($C_1, C_2$) are the tracing variables that were sent to the clients in the first step.
\end{enumerate}

\subsubsection*{\textbf{Step 5: Verification}}

The server needs to verify the group signature associated with message $M$. 

\begin{enumerate}
    \item In first iteration of a session, server collects all public entities, the latest ${C_1,C_2}$ and the $\sigma_{GS}$ available for the particular group from the admin, and computes the verification variables as:
 
\hskip .2 in $\Tilde{B_1} \leftarrow u^{s_ \alpha}C_1^{-c}$, \hskip .3 in
$\Tilde{B_2 }\leftarrow v^{s_ \beta}C_2^{-c}$,

\hskip .2 in $\Tilde{B_3} \leftarrow \hat{e}(C_3,g_2)^{s_x}\hat{e}(h,w)^{- s_ \alpha -s_ \beta}$

\hskip 0.65 in $\hat{e}(h,g_2)^{- s_ {\delta _1} -s_ {\delta _2}} (\hat{e(C_3,w)/e(g_1,g_2))^c}$,

\hskip .2 in $\Tilde{B_4} \leftarrow C_1^{s_x}u^{-s_{\delta _1}}$,
\hskip .2 in
$\Tilde{B_5} \leftarrow C_2^{s_x}v^{-s_{\delta _2}}$.

Then, the server computes the challenger from the received information, denoted as $\Tilde{c}$, as:

\hskip .2 in $\Tilde{c} \leftarrow H(M,C_1,C_2,C_3,\Tilde{B_1},\Tilde{B_2},\Tilde{B_3},\Tilde{B_4},\Tilde{B_5})$.

    \item If $c = \Tilde{c}$, then the signature is verified, and $\sigma_{GS}$ with particular $(C_1, C_2)$ is stored at server side. Otherwise, the server rejects the received message. In this way, the server knows the message $M$ but does not know the signer of the message. The server receives the model updates from all participating members in this way. 

   \item After the first iteration, the server simply fetches the $(\sigma_{GS}, C_1, C_2)$ from its local storage for all the subsequent iterations of the session. It saves the computation and communication cost of the server, and it does not need to compute those keys again.
\end{enumerate}

\subsubsection*{\textbf{Step 6: Federated Aggregation}}

The main task of the server is to compute an aggregated average of the updates received. The formula for local training and gradient aggregation in FedAvg \cite{Brendan} is given as:
\begin{equation}
\label{FedAvg}
\forall~ k, \omega_{t+1}^k \leftarrow \omega_t - \eta g_k,
 \hskip .2 in w_{t+1} \leftarrow \Sigma_{k=1}^K p_k w^k_{t+1}, 
\end{equation}
% where $\eta$ is the fixed learning rate, $K$ is the number of clients, $C$ is the fraction of clients that participate in the process, $\omega$ is the model parameter, $\omega_t$ is the current model, and $\omega_{t+1}$ is the subsequent model. We take $n = \Sigma _kn_k$, where $n$ is the total number of samples in the dataset, which is the same as the number of clients participating in FL, and we set $p_k = \frac{n_k}{n}$. After gradient aggregation, the global model is redistributed to all clients, and the process repeats until a specific condition set by the server is satisfied. 
where $\eta$ is the fixed learning rate, $K$ is the number of clients, $C$ is the fraction of clients that participate in the process, $\omega$ is the model parameter, $\omega_t$ is the current model, and $\omega_{t+1}$ is the subsequent model. We take $n' = \Sigma _kn'_k$, where $n'_k$ is the total number of samples in the dataset for client $k$, and we set $p_k = \frac{n'_k}{n'}$. After gradient aggregation, the global model is redistributed to all clients, and the process repeats until a specific condition set by the server is satisfied.

% regardless of their participation in the previous FL process. In this way, all the clients in the network have the latest updates.

\section{Security Analysis} \label{attack}

\subsection{Security Proof Using AVISPA tool} 

We have provided a formal security verification using AVISPA, which gives the result based on the goal of the algorithm. The algorithm is written in HLPSL code which checks for the authenticity and secrecy of the keys and secret messages. As a proof of concept, we configured to have six entities: server, client 1, client 2, client 3, client 4, and admin. At first, the server sends the global model to all clients in the network. Then, each client sends a group signature request to the admin, to which the admin replies with the necessary keys for the clients. Each client signs its message with the group signature and sends it to the server. The process is iterated. The goal of the algorithm comprises all the secret protocols, majorly named secrecy and authentication, as given below. The simulation code, written in HLPSL, shows that the algorithm can protect secrets between the admin (named A2) and the clients (named C1, C2, C3, and C4), and between the server (named S) and the clients. The proper authentication of all four clients is done, shown by the authentication code.
\begin{lstlisting}
goal
secrecy_of sec_A2_C1,sec_A2_C2,sec_A2_C3,
           sec_A2_C4,sec_S_C1,sec_S_C2,
           sec_S_C3, sec_S_C4  
           
authentication_on  rand1_auth,rand2_auth,
                   rand3_auth,rand4_auth
end goal           
\end{lstlisting}
              
\begin{figure}
    \centering
    
    \includegraphics[width = 7 cm, height = 5.2 cm, frame]{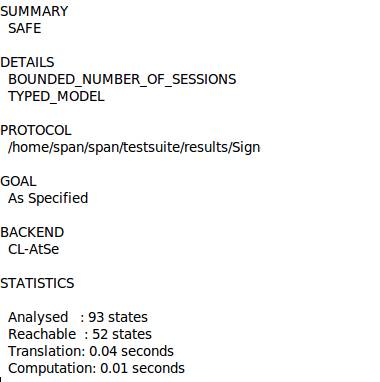}
    \caption{The simulation result: SAFE}
    \label{atse}
      \end{figure}

\begin{figure}
\centering
    \includegraphics[width = 4 cm, height = 3 cm,frame]{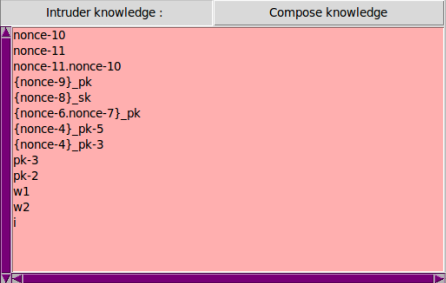}
    \caption{Intruder Knowledge}
    \label{ik}

\end{figure}

The simulation result is given in Figure \ref{atse}, which confirms that our algorithm is safe. Figure \ref{ik} lists all the knowledge acquired by an intruder. It is shown that the attacker can only acquire public keys or encrypted messages.


\begin{table*}[t]
    \begin{center}
    \caption{Protection against various security attacks}
    \label{Tab3}
    %\resizebox{\linewidth}{!}{%
    \begin{tabular}{c|ccccccc}
    \hline
    \multirow{2}{*}{Type of Attack} & \multicolumn{6}{c}{Algorithms} \\ \cline{2-7} 
      & \makecell{Runhua Xu et al.\\\cite{xu2}} &  \makecell{Chai et al.\\\cite{chai}} & \makecell{ Brendan et al.\\\cite{Brendan}} & \makecell{Sun et al.\\\cite{sun}} & \makecell{Xu et al.\\\cite{xu}} & \makecell{GSFL\\(Proposed)} \\ 
    \hline
    1. Honest But Curious Server &\xmark& \xmark &\xmark&\xmark&\xmark&\cmark \\
    2. Inference Attack & \xmark & \xmark & \xmark & \xmark&\xmark&\cmark \\
    3. Selected/Target Clients &\xmark&\xmark&\xmark&\xmark&\xmark&\cmark \\
    % 4. Free Riding Attack  & \xmark & \xmark & \xmark& \xmark&\xmark&\cmark \\
    4. Sybil Attack & \cmark &  \cmark & \cmark&\cmark&\xmark&\cmark \\ 
    5. Denial of Service Attack &\xmark&\xmark&\xmark&\xmark&\xmark&\cmark\\ 
    6. Intruder Creating a Valid Channel & \xmark & \xmark & \cmark &\cmark&\cmark&\cmark \\
     \hline
    \end{tabular}
    %} 
    \end{center}
\end{table*}

\subsection{Security Attacks and Their Prevention}

Security is the main essence of any networking algorithm. Federated learning enhances machine learning through iterative updates without compromising the user's privacy. Hence, the main aim of the algorithm is to learn while protecting personal data. However, like any network technology, FL is also vulnerable to various attacks if not implemented properly. Hence, we have tried to show the possible attacks in the FL network and how our algorithm prevents them from taking place.

\subsubsection{Security Against Honest but Curious Server}

The server cannot be fully trusted when it comes to the privacy of the data. The server might not be interested in the client's data, but it can still peek into some sensitive information if data is not kept securely. An honest but curious server is a server or any other communication member that wants to learn sensitive information about other members from the legitimately received messages.

In our setting, the server receives updates as $(C_i,\sigma_{GS})$ from a list of selected client devices. Since all clients have the same group signature, the server cannot distinguish the particular signer of $\sigma_{GS}$. $C_i$ is the encrypted message with the server's public key. Hence, other members cannot decrypt it. The server can only see the updates, neither raw messages nor the sender of the message. Thus, it makes it difficult to correlate the messages to earlier messages. Therefore, the honest but curious user will not be able to know any sensitive information. 

\subsubsection{Security Against Inference Attacks}

Since the sender of the message is not known even to the server, no entities in the network can relate a message to its previous message. Each message contains only updates and does not contain the entire message. Moreover, raw messages are never sent in the network and are always sent in a processed form. Hence, inferring from previous messages or the same message is impossible.

\subsubsection{Security Against Attack on Selected Clients}

Corrupted clients may join their hands to manipulate the input for the FL process. Hence, the list of clients selected for FL should not be disclosed to other clients. However, clients must be notified about their own selection. We have used encrypted replies from the server-side instead of the open list to ensure this. The key for decryption is only available to the receiver. So, other clients will not be able to know the clients selected for the FL.

% \subsubsection{Security Against Free-Riding Attacks}

% This is usually not an attack but an exploitation of the system. Every honest participant takes part in the FL process. However, some cunning clients benefit from the global model without contributing to the learning process. They can pretend that they have a small number of samples to train, which leads to their rejection from the server-side. \textcolor{red}{[David: This attack seems like it can be applied to other algorithms as well. Unless a unique feature in our proposed protocol explicitly solves this problem, it is safe to remove it (from the text and the table).]}

\subsubsection{Security Against Sybil Attacks}

In a Sybil attack, one user creates multiple fake identities and communicates in the network. This type of attack is challenging, especially in wireless and P2P networks. It is difficult for a receiver to recognize fake identities, leading to other attacks, e.g., data poisoning attacks. Each client is registered with the admin in our model and has a valid group signature. Admin first validates the client's identity, then adds it to the group. It also continuously monitors the suspicious participants, and if any, it immediately revokes them from the group and updates the group signature. It also maintains a revocation table, mentioning the reason for the revocation of any member. 

% If blacklisted, the client can never be added to the group.

\subsubsection{Security Against Intruders Creating Valid Channels} 
An intruder can create a valid channel in-network if it has a valid group signature. It can pretend to be a valid group member if its signature matches the legitimate member's signature. However, we have used strong encryption and hashing techniques to generate the signature, and the elements to generate the signature are not published in the network. Hence, an intruder cannot create a valid channel in the network.

\section{Performance Analysis}\label{perf}

The performance of federated learning is sensitive to the communication and computation costs due to a large number of iterative distributed learning processes on massive resource-constrained client devices. A lower latency will help to achieve higher accuracy.

In this section, we show that the proposed protocol can keep the computation and communication costs low while providing protections against various security attacks, as discussed in Section \ref{attack}. We evaluate the performance of the proposed protocol with existing algorithms in terms of computation, communication, and signaling costs.

% The protocol diagram given in Figure \ref{FLGS} shows the communication involved in our algorithm. The arrows represent the messages exchanged in the FL process. We provide a detailed analysis of the costs in the subsections below.  

\begin{table}[t]
\caption{Unit cost of computing operations}
\label{Tab4}
    \centering
    \begin{tabular} [t]{l c c}
    \hline
    Notation & Operation & Unit Cost (ms) \\ 
    \hline
    $C_{AES}$ & Symmetric encryption & 0.161 \\ 
    $C_{AS}$ & Addition/subtraction & 0.001 \\
    $C_{BP}$ & Bilinear pairing & 4.51 \\
    $C_{DIV}$ & Division & 1.22 \\ 
    $C_{EXP}$ & Exponentiation & 1.0 \\ 
    $C_{H}$ & Hash function & 0.067 \\
    $C_{MOD}$ & Modulus & 1.24 \\
    $C_{MUL}$ & Multiplication & 0.612 \\ 
    $C_{PM}$ & Point multiplication & 1.25 \\  
    $C_{RAND}$& Random number generation & 0.045\\
    $C_{XOR}$ & XOR & 0.002 \\
    % $C_{SCH}$ & Schnorr sign. calculation & 2.004 \\
    \hline
\end{tabular}
\end{table}

% \begin{table}[t]
% \caption{Cost of encryption and decryption for different values of participating clients (in ms)}
% \label{Tab5}
%     \begin{center}%\resizebox{9.0cm}{3.0cm}
%     {
%     \centering
%     \begin{tabular} [t]{p{1.2cm}p{1.2cm}p{1.2cm}p{1.2cm}p{1.2cm}}
%     \hline
%     Algorithm & $n=50$ & $n=100$ & $n=150$ & $n=1000$ \\
%     \hline
%     $C_{enc}$ & 103.78 & 217.85 & 347.89 & 5,210 \\
%     $C_{dec}$ & 55.58 & 102.37 & 145.40 & 1,060 \\
%     \hline
%     \end{tabular}
%     } 
%     \end{center}
% \end{table}

\begin{table*}[t]
\caption{Comparison of average computation cost for $t$ number of iterations ($n = 100$)}
\label{cost2}
    \begin{center}
    {
    \centering
    \begin{tabular} [t]{cc|c|c|c|c|c}
    \hline
    \multirow{2}{*}{Algorithm} & \multirow{2}{*}{Operations} & \multicolumn{4}{c}{Cost of Operations (seconds)} \\ \cline{3-7}
  & &$t=1$ &$t=50$ & $t=100$ & $t=150$ & $t=1000$ \\
    \hline
    Runhua Xu et al. \cite{xu2} & $C_{enc} + C_{dec}+n(C_{enc}+C_{dec}$ + $C_{agg})$ &0.59 &29.45&58.9&88.353&589.025 \\ 
    Chai et al. \cite{chai} & $C_{enc} + C_{dec}+ n( 3(C_{enc}+C_{dec})+ C_{agg})$&1.63 &81.7 &163.4&245&1634 \\  
    Bonawitz et al. \cite{Brendan} & $C_{enc} + C_{dec}+ 9900+ n( C_{enc}+C_{dec} + C_{agg})$ &10.49&524.5&1049&1573.35&10489
    \\ 
    Sun et al. \cite{sun} & $C_{enc} + C_{dec}+ 99+n( C_{enc}+C_{dec} + C_{agg})$&0.68 &34.4&68.8&103.2&688\\
    Xu et al. \cite{xu} & $C_{client-client}+C_{agg}+2C_{enc}+2C_{dec}$ &9.9&495&990&1486&9908
    \\
    Proposed (GSFL) & $C_{admin}+C_{server}+C_{client}$+
    &&&&&
    \\&$(i-1)(C_{enc}+C_{dec}+C_{ver}+C_{agg})$ & 0.08 & 2.36 & 4.69 & 23.3 & 46.6 \\ 
    \hline
    \end{tabular}
    } 
    \end{center}
\end{table*}

\subsection{Computation Cost}

The computation cost is the number of CPU cycles needed to formulate the message. Each operation for computing the various parameters of messages requires a different number of CPU cycles, called the unit cost of those operations. We have taken various unit costs of an operation from \cite{balu}. These unit costs are multiplied by the number of occurrences in the entire process. Table \ref{Tab4} lists the computation cost of operations used in our analysis. 

The computation cost for the entire process can be computed in three parts: (1) computation cost at the admin ($C_{admin}$), (2) computation cost at each client ($C_{client}$), and (3) computation cost at server ($C_{server}$). Let us assume that there are 200 clients in the network, and the server has selected 100 ($n = 100$) of them for the FL process in each iteration. In each simulation session, there are 1000 iterations. The computation cost is directly proportional to the number of iterations involved. 

\subsubsection{Computation Cost at the Admin ($C_{admin}$)}

\begin{itemize}
    \item $C_{admin}$: The admin generates the latest keys necessary to sign the messages, which is basically $A_i, C_1$, and $C_2$. It also generates 2 random number $a$ and $b$. Hence, the total computation cost at the admin is $C_{admin} = C_{DIV} + 3C_{EXP} + 2{C_{RAND}} = 1.22 + 3(1.0) + 2(0.045) = 4.31$ ms.
\end{itemize}
    
\subsubsection{Computation Cost at Each Client ($C_{client}$)}

\begin{itemize}
    \item $C_{client}$: Client uses the keys sent by admin to sign and encrypt the message. Hence, the cost for client can be given as $C_{client} = C_{sig}+ C_{enc}$.
    \item Signature Generation Cost ($C_{sig}$): The signature generation mainly includes computing $C_3$, binding variables, and signing variables computed as
    \begin{equation}\label{eqn1}
    C_{sig} \leftarrow C_{C_3} + C_{BV} + C_{SV} + C_{CH},
    \end{equation} 
    where each term is computed as $C_{C_3} = C_E + C_M + C_{AS} = 1.0 + 0.612 + 0.001 = 1.613$ ms, $C_{BV}$ = $9C_E + 3C_{BP} + 4C_M = 9(1.0) + 3(4.51) + 4(0.612) = 24.978$ ms, $C_{SV} = 5C_{AS}+ 5C_{MUL} = 5(.001) + 5(0.612) = 3.065$ ms, and $C_{CH} = C_H = 0.067$ ms. Putting all above values in (\ref{eqn1}), we get $C_{sig} = 1.613 + 24.978 + 3.065 + 0.067 = 29.723$ ms. 
    \item Encryption Cost ($C_{enc}$): It is the cost of encrypting the message that clients send to the server. The encryption cost can be computed according to \cite{sow} as $C_{enc} = 2C_{EXP} + C_{MUL} =  2(1.0) + 0.612 = 2.612$ ms.
    \item Hence, the total computation cost at the client is $C_{client} = C_{sig}  + C_{enc} = 29.723 + 2.612 = 32.335$ ms.
\end{itemize}

\begin{figure}
\centering

\includegraphics[width = 7.8 cm, height = 4.5 cm]{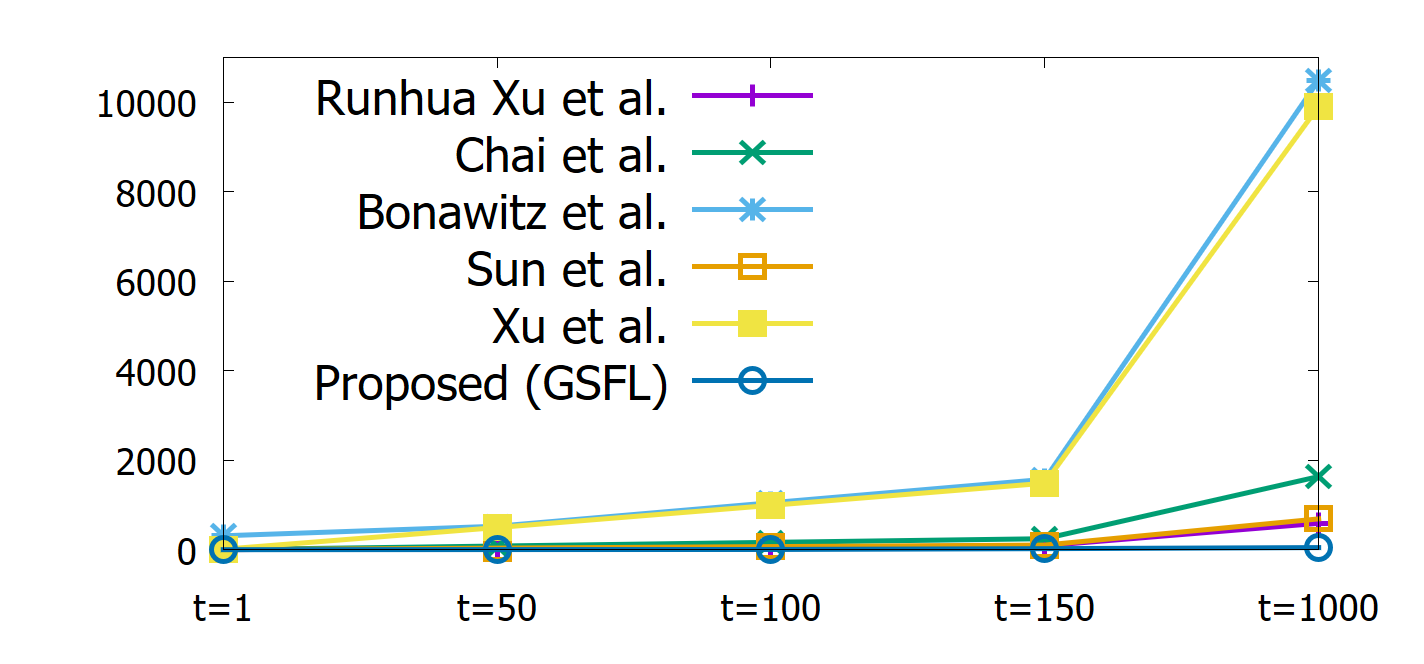}
    \caption{Comparison of computation cost ($n = 100$, $m = 200$)}
    \label{dynacost}

\end{figure}
\subsubsection{Computation Cost at the Server ($C_{server}$)}

\begin{itemize}
    \item Decryption Cost ($C_{dec}$): It is the cost of decrypting the message sent by the client. The decryption cost is computed as $C_{dec} = 2C_{EXP}+ C_{MUL}+ C_{AS} = 2(1.0) + 0.612 + 0.001 = 2.613$ ms. 
    \item Verification Cost ($ C_{ver}$): It is the cost of verifying the signature of a client computed as
    \begin{equation} \label{eqn3}
      C_{ver} \leftarrow  \Tilde{(C_{BV})} + (C_{H}),
    \end{equation}
    where $\Tilde{(C_{BV})} = 8C_M + 12C_E + 5C_{BP} + C_{DIV} = 8(0.612) + 12(1.0) + 5(4.51) + 1.22 = 40.666$ ms.  Putting all above values in (\ref{eqn3}), we get $C_{ver} = 40.666 + 0.067 = 40.733$ ms.
    \item Aggregation Cost ($C_{agg}$): It is the cost needed to aggregate the gradients from the clients participating in the federated learning process. The aggregation cost in FedAvg can be calculated by $\Sigma_{k=1}^K p_k w^k_{t+1}$ as given in (\ref{FedAvg}), where $p_k = \frac{n_k}{n}$.  Hence, total cost of aggregation for n clients would be n multiplications, $n-1$ additions, and $1$ division, where $(K=n)$ is the total number of participating clients.
    $C_{agg_n} = (n-1)(C_{AS})+n(C_{MUL})+C_{DIV} $. Hence aggregation cost for 1 client would be $C_{agg} \approx C_{AS}+C_{MUL} = .001+.612 = .613 ms$
    
    \item Hence, the total computation cost at the server is $C_{server} = C_{dec} + C_{ver} + C_{agg} = 2.613 + 40.733 + 0.613 = 43.959$ ms.
\end{itemize}

Finally, the total computation cost of our algorithm is calculated as $C_{alg}$ =  $C_{server} + C_{client} + C_{admin} = 43.959 + 32.33 + 4.31 = 80.599$ ms.

The above is the cost of computation for the first iteration of FL. However, in the next iterations, we do not need to compute the cost of signature generation. The same signature is used for all the iterations of a session. Hence, we have to add only encryption, decryption, verification and aggregation cost for next iterations, which is $C_{enc}+ C_{dec}+C_{ver}+C_{agg} = 2.612+2.613+40.733+0.613 = 46.571$ ms.

Hence, the computation cost when:
\begin{itemize}
    \item t = 50, $C_{alg}=1\times80.6+49\times46.571 = 2362.579 ms$ 
    \item t = 100, $C_{alg}=1\times80.6+99\times46.571 = 4691.129 ms$ 
     \item t = 150, $C_{alg}=1\times80.6+149\times46.571 = 7019.679 ms$ 
    \item t = 1000, $C_{alg}=1\times80.6+999\times46.571 = 46605.029 ms$ 

\end{itemize}
Similarly, we can compute the computation cost of existing algorithms. The comparison of computation costs is summarized in Table \ref{cost2}. Since the encryption mechanism is not specified in refereed papers, we have assumed that all algorithms have used the ElGamal algorithm for cryptography. Since they have not mentioned any signatures, we have assumed other algorithms have used their digital signature to verify them. In the digital signature, a message is encrypted with the private key of the sender and decrypted at the receiver using the sender's public key. So, we assumed two encryption and two decryption for a single transmission between two clients. The cost of the client-to-client communication needed for secure key transmission is denoted as $C_{client-client}$.  
    
\begin{table}[t]
\caption{Message structure and sizes of fields (in bits)}
\label{Tab6}
\centering
    \begin{tabular}[t]{ccccccccc}
    \hline
    Msg.&$GID$&$MID$&$RAND$&$Payload$&$GS$&$TTL$&$TS$ \\ 
    \hline
    Size & 16 & 16 & 128 & 1024 & 1539 & 8 & 32 \\
    \hline
    \end{tabular}
\end{table}

\begin{figure}
    \includegraphics[width = 7.8 cm, height = 4.5 cm]{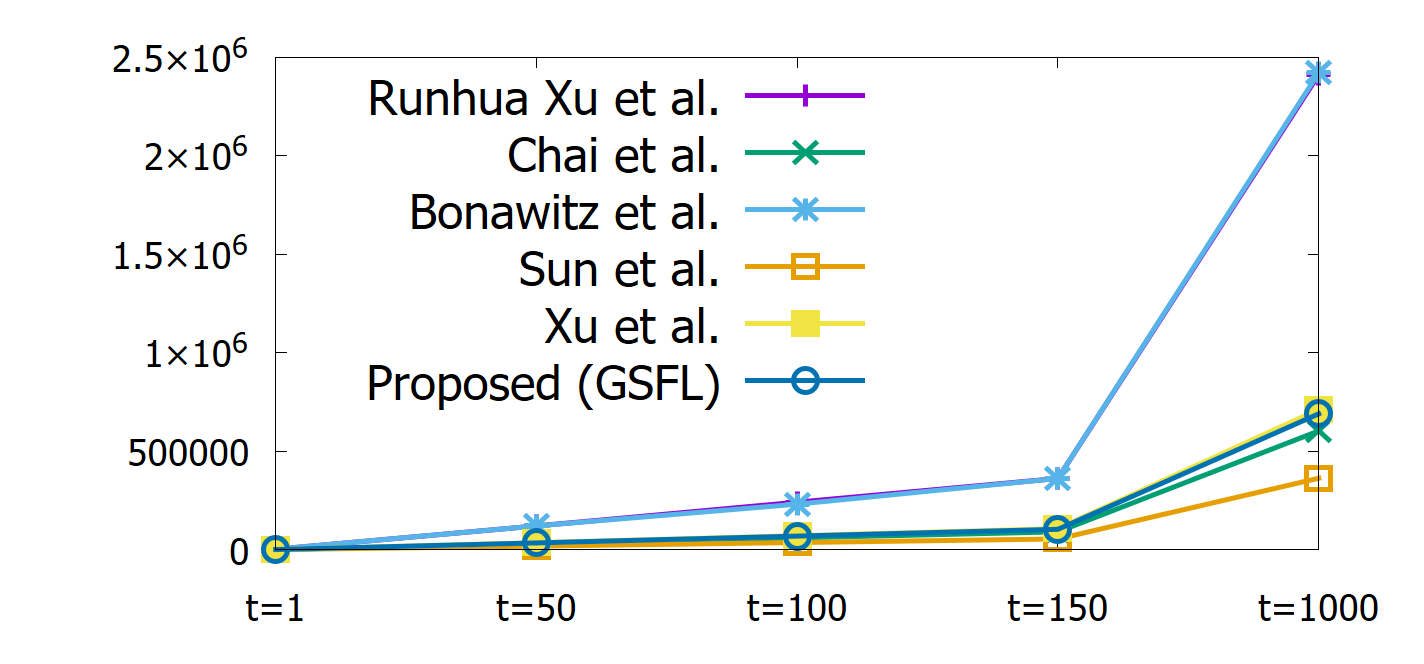}
    \caption{Comparison of communication cost ($n = 100$, $m = 200$)}
    \label{sizeFig}
\end{figure}

% \begin{table}[t]
% \caption{Comparison of communication cost for 1000 number of iterations \textcolor{red}{[David: Need a Table similar to Table IV.]}}
% \label{Tabcost}
%     \begin{center}
%     \centering
%     \resizebox{\linewidth}{!}{
%     \begin{tabular} [t]{c c | c c}
%     \hline
%     Algorithm  & Average Cost &  Algorithm  & Average Cost \\ \hline
%     Runhua Xu et al. \cite{xu2} & 2412 Bytes
%     & Chai et al. \cite{chai}  & 603 Bytes \\
%     %since they are using 300 images per client. Hence, they need to send 300 messages to the server.
%     Brendan et al. \cite{Brendan} & 2412 Bytes
%     & Sun et al. \cite{sun} & 362.7 Bytes \\
%     Xu et al. \cite{xu} & 712 Bytes % MISO is doing 20 iterations, hence multiplied by 20
%     & Proposed (GSFL) & 690 Bytes \\ %\hline
%     \hline
%     \end{tabular}
%     }
%     \end{center}
% \end{table}

\begin{table*}[t]
\caption{Comparison of average communication cost for $t$ number of iterations ($n = 100$)}
\label{comm}
    \begin{center}
    {
    \centering
    \begin{tabular} [t]{c |c| c |c| c| c}
    \hline
    \multirow{2}{*}{Algorithm} &  \multicolumn{4}{c}{size of messages (in bytes)} \\ \cline{2-6}
   &$t=1$ &$t=50$ & $t=100$ & $t=150$ & $t=1000$ \\
    \hline
    Runhua Xu et al. \cite{xu2}  &2412 &120600&241200&361800&2412000 \\ 
    Chai et al. \cite{chai} &603 &30150 &60300&90450&603000 \\  
    Bonawitz et al. \cite{Brendan}  &2412&120600&241200&361800&2412000
    \\ 
    Sun et al. \cite{sun} &363&18150&36300&54450&363000\\
    Xu et al. \cite{xu}  &712&35600&71200&106800&712000
    \\
    Proposed (GSFL) &690&34500&69000&103500&690000 \\ 
    \hline
    \end{tabular}
    } 
    \end{center}
\end{table*}

\begin{table*}[t]
\caption{Comparison of average signaling cost for $t$ number of iterations ($n = 100$)}
\label{Tabsig}
    \begin{center}
    {
    \centering
    \begin{tabular} [t]{c c |c| c |c| c| c}
    \hline
    \multirow{2}{*}{Algorithm} & \multirow{2}{*}{Operations} & \multicolumn{4}{c}{Cost of Operations (seconds)} \\ \cline{3-7}
  & &$t=1$ &$t=50$ & $t=100$ & $t=150$ & $t=1000$ \\
    \hline
    Runhua Xu et al. \cite{xu2} & $m+m+m+t\times(n+n+n+1)+m$&1,101 &15,850 &30,900&45,950&301,800 \\ 
    Chai et al. \cite{chai} & $m+m+t\times(n+n+1+n+n)+m$&1,001 &20,650 &40,700&60,750&401,600 \\  
    Bonawitz et al. \cite{Brendan} & $m+m+t\times(n+n(n-1)+n+1)+m$ &10,701&505,650&1,010,700&1,515,750&10,101,600
    \\ 
    Sun et al. \cite{sun} & $m+t\times(n(n-1)/100+n)+m$&1,490 &10,350&20,300&30,250&199,400\\
    Xu et al. \cite{xu} & $m+m+t\times(n+n(n-1)+n+1)+m$ &10,701&505,650&1,010,700&151,5750&10,101,600
    \\
    Proposed (GSFL) & $m+m+m+t\times(n+n+1)+m$ & 1,001 & 10,850 & 20,900 & 30,950 & 201,800 \\ 
    \hline
    \end{tabular}
    } 
    \end{center}
\end{table*}

\begin{figure}
    \centering
    \includegraphics[width = 7.8 cm, height = 4.5 cm]{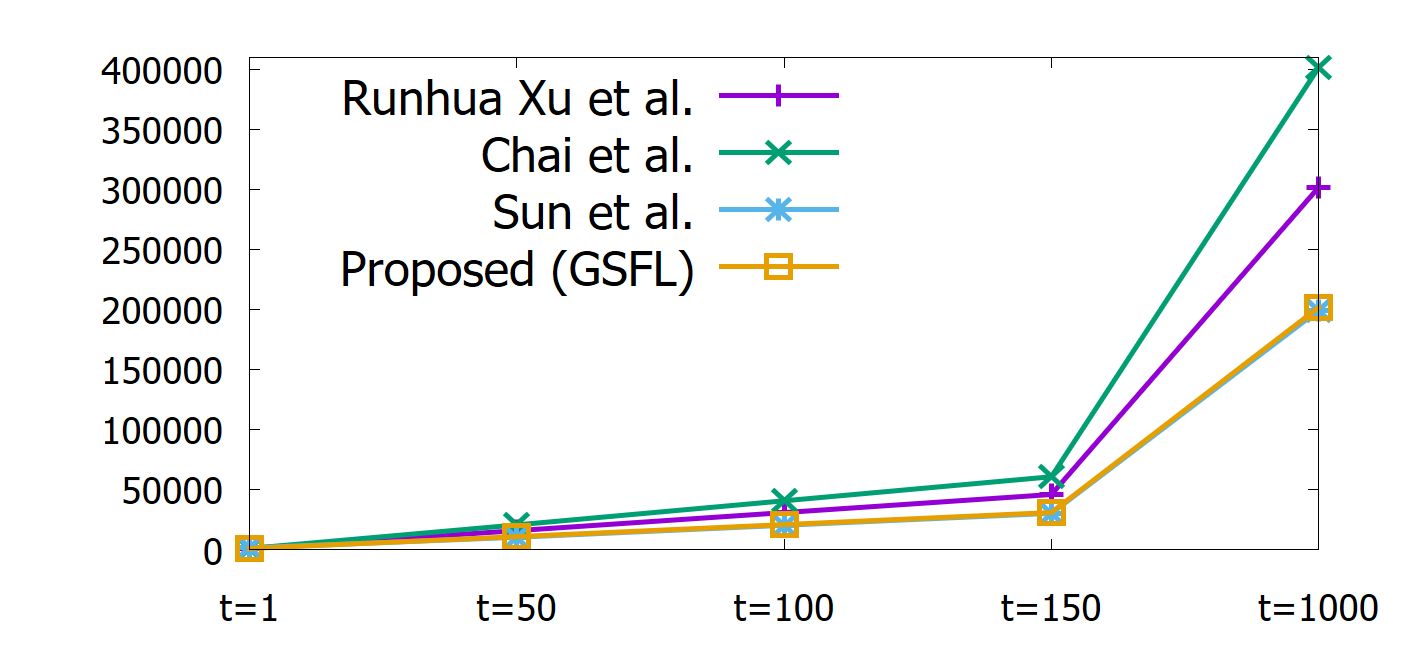}
    \caption{Comparison of signalling cost ($n = 100$, $m = 200$)}
    \label{signal}
\end{figure}

\subsection{Communication Cost}

This section describes the cost caused by the size of the packet transmitted from clients and the admin to the server. Note that the packets sent from the server to the clients have no specific change, which will be the same as other traditional approaches. Hence, we are not comparing the packet size from the server to clients. 

For our algorithm, the packet size can be given by the final output message we are sending in the network, which is $(c_i,\sigma_{GS})$. The size of $c_i$ is same for all algorithms, suppose it is 1024-bits. On the other hand, signature varies depending on the algorithms. In our case, it is given as $\sigma_{GS} \leftarrow (C_1,C_2,C_3,c,s_\alpha,s_\beta,s_x,s_{\delta _1},s_{\delta _2})$.  Each variable in a signature is 171-bits long, and there are nine variables in the signature. So, the total signature size is $9\times171 = 1539$-bits. The message structure is given in Table \ref{Tab6}. The total size of the message per communication is the sum of all fields calculated as $M_{size} = GID_{Size} + MID_{Size} + RAND_{Size} + Payload_{Size} +GS_{Size} + TTL_{Size} + TS_{Size} = 16+16+128+1024+1539+8+32 = 2763$ $bits = 345$ bytes. Note that the client sends a message twice to the server: (1) request message for FL training and (2) actual data. Therefore, the total message size from the client will be 690 bytes.

In the algorithms that use personal signatures (not a group signature), each client sends a double encrypted message to the server, one for encryption and another for signature. Therefore, there is no separate section for signature in the final packet, and the size is saved. However, clients may need to send a separate message for further privacy preservation. For comparison, the message size of other algorithms is obtained as described below:
 \begin{enumerate}
     \item In Bonawitz et al. \cite{Brendan}, each client sends two messages to the server. However, it also sends $n-1$ messages to other clients participating in the FL. Suppose there are $n=100$ clients participating in FL. The total packet size for all communication in one iteration of FL would be $M_{size_{Bonawitz}} =  MID_{Size} +RAND_{Size}+ Payload_{Size} + TTL_{Size} + TS_{Size} = 16+128+1024+8+32 = 1080$ $bits = 135$ bytes. In addition to the server, each participating client needs to communicate with other $n-1$ clients. These messages contain all other fields as above messages other than the payload, which makes it 184-bits or 23 Bytes. So, the total size of all messages will be $(n-1)\times 23+M_{size_{Bonawitz}} $, which is $(99\times23+135) = 2412$ bytes.
     \item In Sun et al. \cite{sun}, we have considered the clients communicate with each other after ten iterations of FL. Hence, the total bandwidth after every ten iterations would be the same as \cite{Brendan}. However, no communication takes place for the next nine training iterations, and the bandwidth is saved for nine iterations. So, computing the average bandwidth for 1000 iteration will result in $(1000\times135+ 99\times100\times23)/1000 = 362.7$ bytes.
     \item Similarly, the communication cost for other existing algorithms has been computed, and the result is given in Table \ref{comm}.
\end{enumerate}
 
\subsection{Signalling Cost}

This section compares the total number of message transmissions between client and server in a session. For handshaking and exchanging authentication information, the client and server exchange a few messages which are not part of actual updates. Those transmissions are costs of the network and need to be minimized. The timing diagram for our algorithm is given in Figure \ref{signal}. 

We can compute that the total number of signals transmitted for an iteration of FL is $m+m+m+n+n+1+m$, where $m$ is the total number of clients in the network, and $n$ is the total number of clients chosen for the FL. We assumed 2000 clients in the network where half of them are chosen by the server to participate in FL (i.e., $m=2000$ and $n=1000$). In our algorithm, the signalling cost is $2000+2000+2000+1000+1000+1+2000 = 10001$. On the contrary, the signaling cost in \cite{Brendan} would be the highest among all algorithms we have referred to. Other steps would be the same, but in place of admin messages to the client, there are $n(n-1)$ messages for the client-to-client communication, resulting in $2000+2000+1000+1000(1000-1)+1000+1+2000 = 1007001$ signals. Similarly, we can compute the signaling cost of other algorithms, and the output is given in Table \ref{Tabsig}.

\section{Conclusion} \label{con}

We proposed a new group signature protocol to handle clients' privacy preservation in federated learning. The proposed protocol efficiently integrates the traditional group signature with federated learning by considering the iterative learning process. Moreover, it does not contradict the architecture and assumptions of the traditional federated learning approach. We provided a mathematical analysis of the computation, communication, and signaling cost, which are closely associated with the performance of federated learning algorithms. Also, we have provided a comprehensive security analysis and proved the safety of the proposed protocol using a formal verification tool. Our protocol outperforms existing algorithms in terms of efficiency and handles various security attacks in the federated learning environment. There are still many challenges that remain. One of our future work is to handle dynamic changes in group members. We are also working on client selection of the federated learning process while the identity of the clients is hidden from the server.

\section*{Acknowledgement}

This research was supported by the MSIT, Korea, under the National Research Foundation (NRF), Korea (2022R1A2C4001270), and the ITRC support program (IITP-2020-2020-0-01602) supervised by the IITP.

% \ifCLASSOPTIONcaptionsoff
%   \newpage
% \fi
\bibliographystyle{elsarticle-num}

\begin{thebibliography} {99}
\bibitem{chris} 
Chris O'Brien. https://venturebeat.com/2021/03/29/canalys-more-data-breaches-in-2020-than-previous-15-years-despite-10-growth-in-cybersecurity-spending 2021 
\bibitem{hu} 
Hu, Hongsheng, et al. "Source Inference Attacks in Federated Learning." 2021 IEEE International Conference on Data Mining (ICDM). IEEE, Auckland, New Zealand, Dec. 2021.
\bibitem{aslam}
Aslam, Sidra, Aleksandar Tošić, and Michael Mrissa. "Secure and Privacy-Aware Blockchain Design: Requirements, Challenges and Solutions." Journal of Cybersecurity and Privacy vol.1, no.1, pp.164-194, Mar. 2021.
\bibitem{Bonawitz}
Bonawitz, Keith, et al. "Practical secure aggregation for privacy-preserving machine learning." proceedings of the 2017 ACM SIGSAC Conference on Computer and Communications Security. Dallas, Texas, USA, pp.1175-1191, Oct. 2017.
% \bibitem{Bon}
% Bonawitz, Keith, et al. "Practical secure aggregation for federated learning on user-held data." arXiv preprint arXiv:1611.04482, Nov. 2016.
\bibitem{yong}
Li, Yong, et al. "Privacy-preserving federated learning framework based on chained secure multiparty computing." IEEE Internet of Things Journal, vol.8, no.8, pp.6178-6186, Apr. 2021.

\bibitem{Ang}
Li, Ang, et al. "FedMask: Joint Computation and Communication-Efficient Personalized Federated Learning via Heterogeneous Masking." Proceedings of the 19th ACM Conference on Embedded Networked Sensor Systems, New York, NY, USA, pp.42-55, Nov. 2021.
\bibitem{wan}
Wan, Xicheng, et al. "Towards Privacy-Preserving and Verifiable Federated Matrix Factorization." arXiv preprint arXiv:2204.01601, Apr. 2022.
\bibitem{Hartmann}
Hartmann, Florian. \emph{Federated Learning} línea]. Available: https://florian. github. io/federated-learning/, \relax [Último acceso: 15 10 2019], May. 2018.

\bibitem{kim}

Kim, Hyesung, et al. \emph{Blockchained on-device federated learning.} IEEE Communications Letters, vol.24, no.6, pp.1279-1283, Jun. 2020

\bibitem{Zhu}
Zhu, Saide, et al. "Secure verifiable aggregation for blockchain-based federated averaging." High-Confidence Computing vol.2, no.1,  pp.100046, Mar. 2022.
\bibitem{yijing}
Li, Yijing, et al. "Privacy-preserved federated learning for autonomous driving." IEEE Transactions on Intelligent Transportation Systems, pp. 1-12,Jun. 2021.

\bibitem{Wei}
Wei, Kang, et al. "Federated learning with differential privacy: Algorithms and performance analysis." IEEE Transactions on Information Forensics and Security vol.15, pp.3454-3469, Apr. 2020.
\bibitem{AVISPA}
Armando, Alessandro, et al. "The AVISPA tool for the automated validation of internet security protocols and applications." International conference on computer aided verification. Springer, Berlin, Heidelberg, pp.135-165, Jul. 2005.

\bibitem{Timothy}
Yang, Timothy, et al. "Applied federated learning: Improving google keyboard query suggestions." arXiv preprint arXiv:1812.02903 2018.

\bibitem{manoj}
Prabhakaran, Manoj, and Mike Rosulek. "Reconciling Non-malleability with Homomorphic Encryption." Journal of Cryptology vol 30, no.3  pp. 601-671.2017


\bibitem{Brendan}
McMahan, Brendan, et al. \emph{Communication-efficient learning of deep networks from decentralized data} Artificial Intelligence and Statistics. \relax PMLR, vol.54, pp.1273-1282, Apr. 2017.
\bibitem{david}
Chaum, David, and Eugène van Heyst. "Group signatures." Workshop on the Theory and Application of of Cryptographic Techniques. Springer, Berlin, Heidelberg, vol.547, pp.257-265, May. 2001.
\bibitem{sneha}
Kanchan, Sneha, and Bong Jun Choi. "Group Signature Based Federated Learning Approach for Privacy Preservation." 2021 International Conference on Electrical, Computer and Energy Technologies (ICECET). IEEE, Cape Town, South Africa, Dec. 2021.
\bibitem{gsis}
Lin, Xiaodong, et al. "GSIS: A secure and privacy-preserving protocol for vehicular communications." IEEE Transactions on vehicular technology, vol.56, no.6, pp.3442-3456, Dec. 2007.
\bibitem{Boneh}
Boneh, Dan, Xavier Boyen, and Hovav Shacham, \emph{Short group signatures} Annual international cryptology conference, \relax Springer, Berlin, Heidelberg, vol.3152, pp.41-45, Aug. 2004.
\bibitem{moti}
Tsiounis, Yiannis, and Moti Yung. "On the security of ElGamal based encryption." International Workshop on Public Key Cryptography. Springer, Berlin, Heidelberg, pp.117-134, Jan. 1998.

\bibitem{wiki}
"ElGamal encryption."Wikipedia, Wikimedia Foundation, 24 Feb. 2019. https://en.wikipedia.org/wiki/ElGamal\_encryption
\bibitem{balu}
Parne, Balu L., Shubham Gupta, and Narendra S. Chaudhari \emph{ Segb: Security enhanced group based aka protocol for m2m communication in an iot enabled lte/lte-a network} IEEE Access \relax vol.6, pp.3668-3684, Jan. 2018.



\bibitem{sow}
Sow, Demba, and Pascal Lafourcade. "Linear generalized ElGamal encryption scheme." Cryptology ePrint Archive. Paris, France, Jul. 2020.


\bibitem{xu2} 
Xu, Runhua, et al. \emph{FedV: Privacy-Preserving Federated Learning over Vertically Partitioned Data.} arXiv preprint \relax arXiv:2103.03918, Jun. 2021.


\bibitem{chai}
Chai, Di, et al.\emph{FedEval: A Benchmark System with a Comprehensive Evaluation Model for Federated Learning} arXiv preprint \relax arXiv:2011.09655, Nov. 2020.

\bibitem{sun}
Sun, Tao, Dongsheng Li, and Bao Wang. \emph{Decentralized Federated Averaging}  arXiv preprint \relax arXiv:2104.11375, Apr. 2021.


\bibitem{xu}
Xu, Chunmei, et al. \emph{Learning Rate Optimization for Federated Learning Exploiting Over-the-air Computation} arXiv preprint \relax arXiv:2102.02946, Apr. 2021.

\end{thebibliography}

\end{document}